\documentclass[preprint]{aastex}
\usepackage{amsmath}
\usepackage[displaymath, mathlines]{lineno}

\slugcomment{To appear in the 2015 March 20 issue of ApJ Letters}
\shorttitle{Rapid Spin and Trailing Fragments of \mbox{P/2012 F5}}
\shortauthors{Drahus et al.}

\begin{document}

\title{Fast Rotation and Trailing Fragments of the Active Asteroid \\ \mbox{P/2012 F5} (Gibbs)\altaffilmark{*}}

\author{Micha{\l} Drahus}
\affil{Astronomical Observatory, Jagiellonian University, Krak{\'o}w, Poland}
\email{drahus@oa.uj.edu.pl}

\author{Wac{\l}aw Waniak}
\affil{Astronomical Observatory, Jagiellonian University, Krak{\'o}w, Poland}

\author{Shriharsh Tendulkar}
\affil{Division of Physics, Mathematics \& Astronomy, California Institute of Technology, \\ Pasadena, CA, USA}

\author{Jessica Agarwal}
\affil{Max Planck Institute for Solar System Research, G{\"o}ttingen, Germany}

\author{David Jewitt}
\affil{Department of Earth, Planetary \& Space Sciences and Department of Physics \& Astronomy, \\ University of California at Los Angeles, Los Angeles, CA, USA}

\and

\author{Scott S. Sheppard}
\affil{Department of Terrestrial Magnetism, Carnegie Institution for Science, Washington, DC, USA}

\altaffiltext{*}{The data presented herein were obtained at the W. M. Keck Observatory, which is operated as a scientific partnership among the California Institute of Technology, the University of California and the National Aeronautics and Space Administration. The Observatory was made possible by the generous financial support of the W. M. Keck Foundation.} 

\begin{abstract}
While having a comet-like appearance, \mbox{P/2012 F5} (Gibbs) has an orbit native to the Main Asteroid Belt, and physically is a km-sized asteroid which recently (mid 2011) experienced an impulsive mass ejection event. Here we report new observations of this object obtained with the Keck~II telescope on UT~2014 August~26. The data show previously undetected \mbox{200-m} scale fragments of the main nucleus, and reveal a rapid nucleus spin with a rotation period of $3.24 \pm 0.01$~hr. The existence of large fragments and the fast nucleus spin are both consistent with rotational instability and partial disruption of the object. To date, many fast rotators have been identified among the minor bodies, which, however, do not eject detectable fragments at the present-day epoch, and also fragmentation events have been observed, but with no rotation period measured. \mbox{P/2012 F5} is unique in that for the first time we detected fragments and quantified the rotation rate of one and the same object. The rapid spin rate of \mbox{P/2012 F5} is very close to the spin rates of two other active asteroids in the Main Belt, 133P/Elst-Pizarro and (62412), confirming the existence of a population of fast rotators among these objects. But while \mbox{P/2012 F5} shows impulsive ejection of dust and fragments, the mass loss from 133P is prolonged and recurrent. We believe that these two types of activity observed in the rapidly rotating active asteroids have a common origin in the rotational instability of the nucleus.
\end{abstract}


\section{Introduction}\label{Sec_Intro}

In the classical picture of the solar system, asteroids and comets were regarded as physically distinct classes of object, the former having been made of rocks and the latter recognized as ``dirty snowballs''. This view was consistent with the fact that near the Sun cometary ices sublimate producing the familiar comae and tails, while asteroids remain inactive. The two classes were also dynamically different, comets having been formed in the giant-planet region and stored at the periphery of the solar system, while asteroids originating in the terrestrial planet region where they continue to orbit. However, groundbreaking discoveries in the last decade \citep[e.g.][]{Hsi06} show that this picture --- although roughly valid --- is no longer that simple, as one in $10,000$ or $15,000$ main-belt asteroids displays the characteristic cometary activity \citep[e.g.][]{Hsi15,She15}.

From the total number of 18 active asteroids identified to date (including the so-called Main Belt Comets), some show protracted mass loss (in four cases confirmed to be recurrent), commonly explained by weak sublimation of H$_2$O in the outer Main Belt, or thermal disintegration very close to the Sun \citep[e.g.][]{Jew15}. However, the mass loss of some other active asteroids in the Main Belt appears to be generated by episodic dust ejection of a very short duration, which clearly cannot be explained by a slow and steady process such as ice sublimation. Instead, both hypervelocity impact and rotational instability (the latter resulting from YORP- or sublimation-driven spin-up) have been regarded as possible mechanisms \citep[e.g.][]{Jew12}. This particularly interesting group of active asteroids is currently formed by the large (596)~Scheila \citep[e.g.][]{Bod11,Jew11}, and the small: \mbox{P/2010 A2} \citep[e.g.][]{Jew10a,Sno10}, \mbox{P/2012 F5} \citep{Ste12,Mor12}, \mbox{P/2013 P5} \citep[e.g.][]{Jew13b}, and \mbox{P/2013 R3} \citep{Jew14a}. Except for the slowly rotating (596)~Scheila which was evidently impacted \citep[e.g.][]{Ish11a,Bod14}, mass-shedding or splitting resulting from rotational instability have been identified as likely explanations for the activity of the smaller objects, but the critical piece of evidence --- namely the measurement of the rotation period --- have been missing up until now.

We also observe a growing number of rapidly rotating asteroids, which experience centrifugal forces greater than the centripetal forces from self-gravity \citep[e.g.][]{Kwi10}. Given that these objects do not show any traces of recent mass loss, they must be protected from rotational disruption by an additional component of the centripetal force, generated by a non-zero material strength. For a long time it has been hypothesized that such fast rotators can only be monolithic asteroids smaller than a couple hundred meters in size, but recent discoveries \citep[e.g.][]{Roz14,Cha14} show that in exceptional cases also larger objects, supposedly primitive rubble piles, can tolerate spin rates beyond the limit of stability ensured by the sole self-gravity.

\section{Observations}\label{Sec_Obs}

In order to test the hypothesis of rotational instability being the trigger of catastrophic mass loss from small solar-system bodies, we obtained detailed data of \mbox{P/2012 F5} (Gibbs), a main belt asteroid previously identified to have ejected dust in a single short-duration event in mid 2011 \citep{Ste12,Mor12}. The object has a dynamically stable orbit \citep{Ste12} of a nearly-circular shape, stretching between $2.88$ and $3.13$~AU from the Sun. The location of \mbox{P/2012 F5} in the outer Main Belt as well as the neutral optical colors \citep{Ste12,Nov14} are consistent with \mbox{C-type} asteroids. Given that the nucleus appeared star-like already in May 2013, with an equivalent diameter equal to about 2~km \citep{Nov14}, the selected target created an excellent opportunity for a robust measurement of the nucleus rotation period, but also for a deep search for fragments --- the two key diagnostics of rotational disruption.

The data were taken on UT~2014 August~26 using the 10.0~m Keck~II telescope atop Mauna Kea, Hawaii. We observed with DEIMOS \citep{Fab03} in imaging mode, collecting a time series of $92\times180$ and $7\times45$~sec exposures. All images were taken in the R band with the image scale of $0.1185$~arcsec per pixel. The sky was dark and photometric, and the point spread function (PSF) varied in full width at half maximum (FWHM) from $0.74$ to $1.29$~arcsec, with a median equal to $0.90$~arcsec. At the epoch of observation \mbox{P/2012 F5} was located $2.938$~AU from the Sun and $1.946$~AU from the Earth, the geocentric phase angle was $4.65^\circ$, and the orbit plane angle was $-4.5^\circ$.

\section{Data Reduction}\label{Sec_Red}

The images were corrected for the basic instrumental effects using master bias and twilight flat-field frames obtained on the night of the observations. Internal photometric calibration was secured by selection of 20 reference stars (more specifically, star-like objects) present in the field of view. Absolute calibration was done through observations of two standard stars \citep{Lan92} in the evening twilight (airmass of $\sim1.1$ and $\sim1.9$). The absolute calibration has a standard error of $0.014$~mag, stemming from the photometric uncertainty of one of the standard stars, while the relative photometric offsets of the individual frames are known to a few millimagnitude precision.

Brightness was measured in a variable-size photometric aperture controlled by the size of the PSF. Treating the PSF as a two-dimensional probability density function, we defined its half-width through the shape-independent standard deviation, measured along the short axes of the reference stars (elongated by non-sidereal tracking). The aperture radius selected for the photometry of \mbox{P/2012 F5} was three times the PSF half-width, while the radius employed for the reference and standard stars was five times the PSF half-width. The photometric errors were estimated individually for each data point. They are dominated by the signal-to-noise ratio and the uncertainty of the PSF size, and were minimized by the selected aperture scaling factors. (Note that the systematic error introduced by the uncertainty of the absolute calibration is not applicable to the individual data points in this case.) Faint background stars or galaxies that would affect the photometry by more than $0.01$~mag were identified and subtracted using a simple procedure. First, we selected four frames exposed around the same time as the affected frame, but with the unwanted object sufficiently separated from the nucleus of \mbox{P/2012 F5}. These four frames were photometrically calibrated (through the reference stars), aligned and averaged, and the result of these operations subtracted from the affected frame. The uncertainty introduced by this procedure was included in the photometric errors. The photometric errors were later propagated into the uncertainties of the light curve parameters (period, mean level, range of variation), using a Monte Carlo method with 5000 simulated data sets. The resulting error of the mean level was additionally combined with the much larger error of the absolute calibration.

Besides retrieving photometric information from the individual frames, we also combined all the data to produce a single ultra-deep image. The long and short exposures taken at varying airmass were photometrically equalized (through the reference stars) and averaged with the same weights. The average image was computed in two versions: one aligned on the nucleus of \mbox{P/2012 F5} and the other one aligned on the reference stars. In the computation of the former, we identified and rejected the pixels occupied by distant stars and galaxies in the individual frames. Thanks to this procedure, the resulting average image features a remarkably clean background. 

\section{Results}\label{Sec_Results}

\begin{figure}
\epsscale{0.60}
\plotone{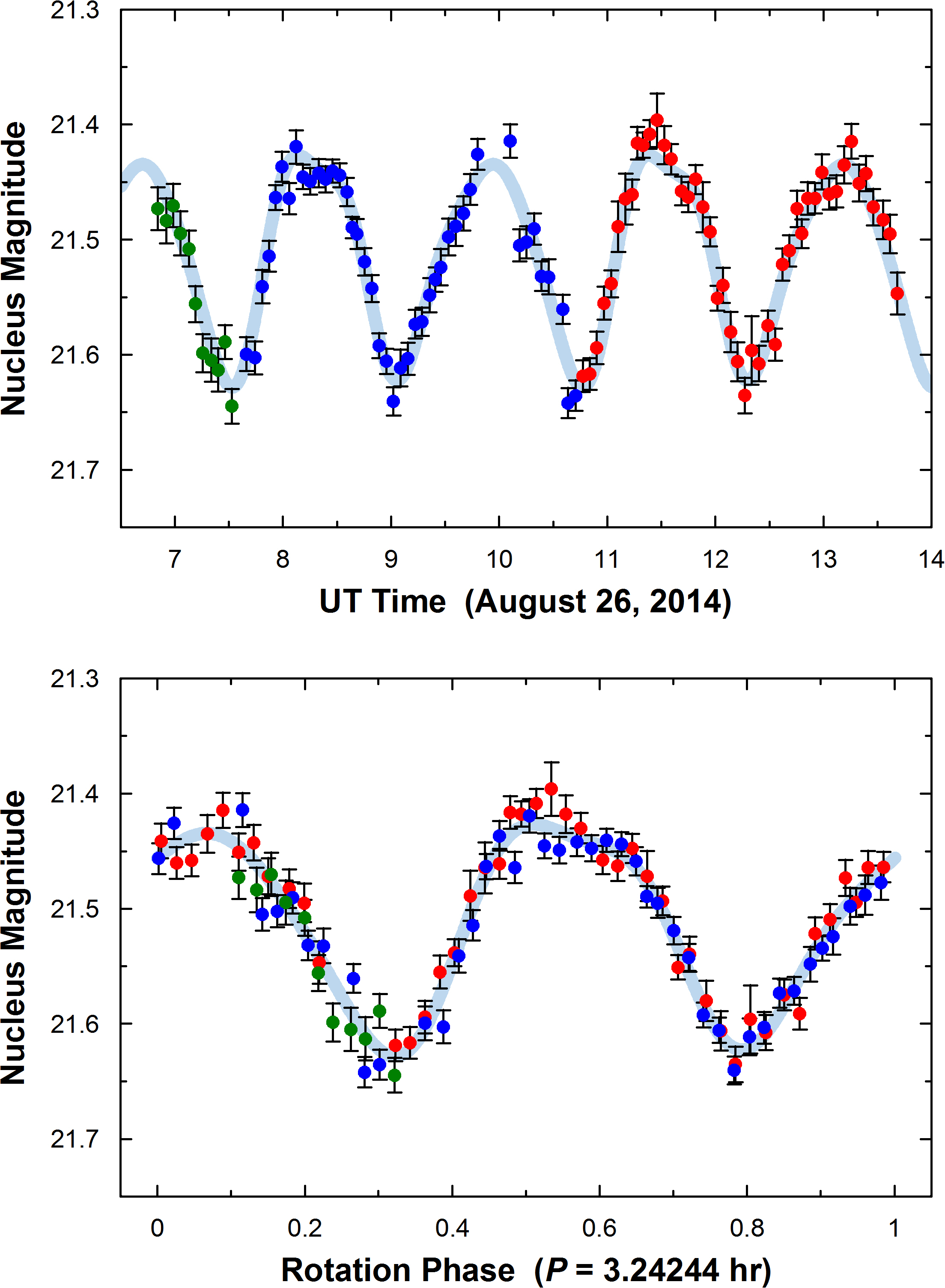}
\caption{\mbox{R-band} magnitude of the nucleus of \mbox{P/2012 F5} presented as a function of time (\emph{top}) and as a function of the nucleus rotation phase (\emph{bottom}). Subsequent rotation cycles are coded by different colors. The underlying light thick line is the best fit of a sum of five harmonics, which optimally represents the data, and which provided the nucleus rotation period, average brightness, and range of brightness variation.\label{Fig_LightCurve}}
\end{figure}

Photometric analysis of the data reveals a double-peaked light curve (Fig.~\ref{Fig_LightCurve}), naturally produced by a rotating elongated object reflecting different amounts of sunlight during the diurnal cycle. The observed periodicity in the light curve is consistent with a nucleus rotation period of $3.24\pm0.01$~hr, which we determined using standard periodicity-search algorithms \citep{Dra06}. The data are represented by a fitted function (Fig.~\ref{Fig_LightCurve}), which gives the average nucleus brightness of $21.513\pm0.014$~mag and the range of brightness variation of $0.201\pm0.006$~mag. Assuming the asteroidal photometric phase function \citep{Bow89} with $G = 0.15$ characteristic of the \mbox{C-types} \citep{Luu89}, we obtain the absolute average brightness $H_\mathrm{R} = 17.325\pm0.014$~mag, corresponding to the equivalent nucleus diameter of $1.77\pm0.01$~km if the geometric albedo is 5\%. Additionally, the measured range of brightness variation places a robust lower limit on the long-to-short axis ratio of a prolate spheroid equal to $1.204\pm0.007$. The nucleus diameter is consistent with the value $\sim2$~km reported by \citet{Nov14}, confirming that the dust contamination was negligible already in May 2013. It is also consistent with two independent upper limits of $\sim6$ and $\sim4$~km reported by \citet{Ste12}, but disagrees with the diameter of $200$ to $300$~m obtained indirectly by \citet{Mor12} from a model of the dust trail.

\begin{figure}
\epsscale{1.00}
\plotone{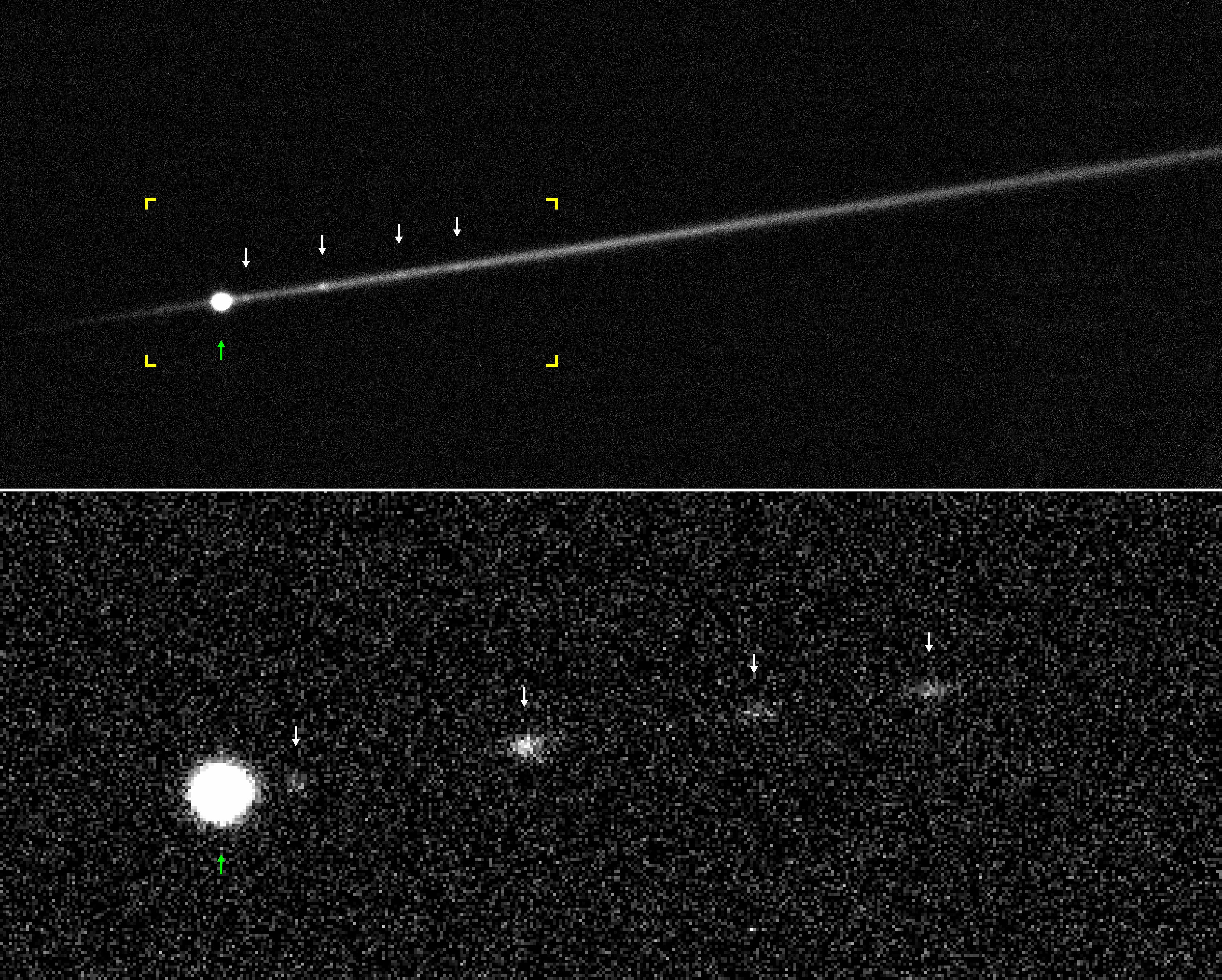}
\caption{Average \mbox{R-band} image of \mbox{P/2012 F5}. \emph{Top panel} subtends $2.5 \times 1.0$~arcmin and shows a star-like nucleus (upwards arrow) and four fragments of the nucleus (downwards arrows) embedded in a dust trail. The solid corners indicate the region enlarged in the bottom image. \emph{Bottom panel} shows a close-up view of the main nucleus and fragments obtained upon numerically removing the dust trail and a small portion of the nucleus ``wing'' (cf. Fig.~\ref{Fig_TrailProfiles}). Presented with a $3\times$ higher spatial resolution and a $2\times$ higher contrast compared to the top image. The trail-subtracted image enabled us to accurately measure the brightness of the fragments with respect to the average brightness of the nucleus.\label{Fig_AvgImage}}
\end{figure}

\begin{figure}
\epsscale{0.60}
\plotone{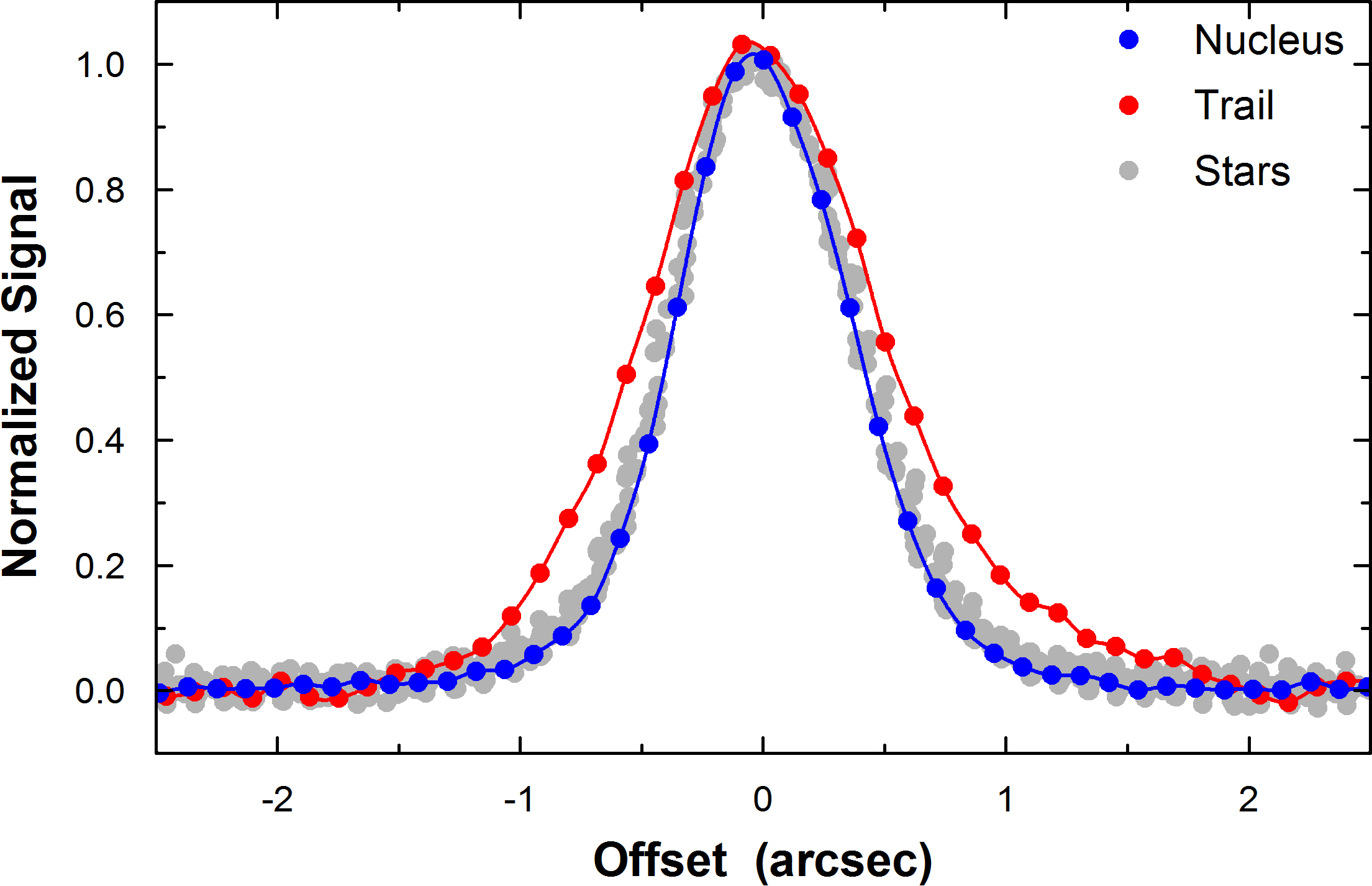}
\caption{\mbox{R-band} brightness profiles of the nucleus, trail, and $15$ field stars (a subset of the best reference stars selected for photometry). We sampled the profiles in the averaged data aligned either on stars --- for the measurement of the stellar profiles, or on the nucleus --- for the measurements of the nucleus (in the trail-subtracted image) and trail (original image). The profiles of the stars were sampled perpendicular to the direction of elongation (caused by non-sidereal tracking) through the brightest pixel. The profile of the trail was measured perpendicular to the axis of the trail in the region of maximum surface brightness, but uncontaminated by the detected nucleus fragments (cf. Fig.~\ref{Fig_TrailProfiles}). In order to improve the signal-to-noise ratio, each point of the trail profile is an average from $80$ pixels extending over $9.48$~arcsec (in one piece) along the trail. The profile of the nucleus was measured in the same way as the profiles of the stars.\label{Fig_Profiles}}
\end{figure}

\begin{figure}
\epsscale{0.60}
\plotone{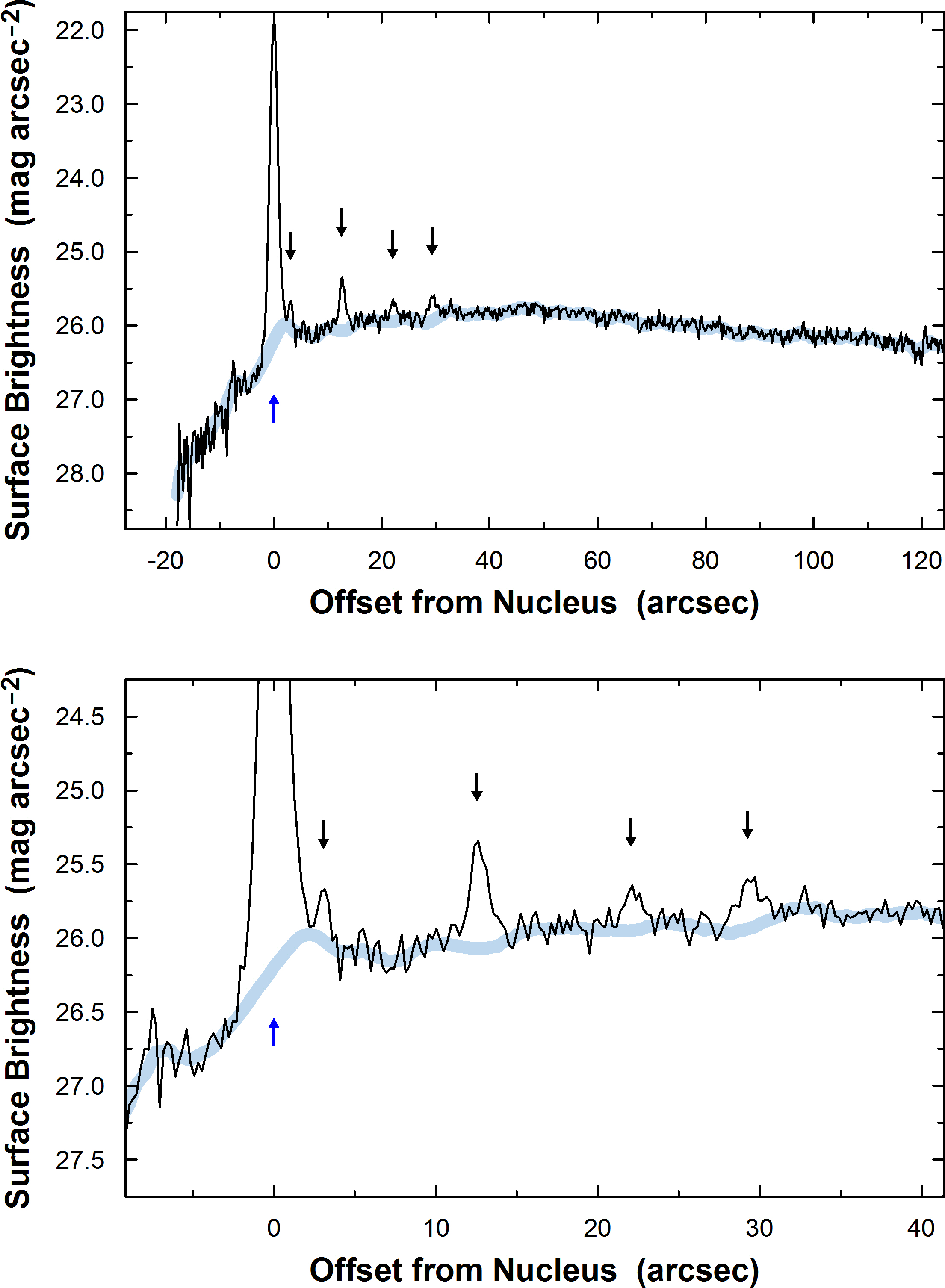}
\caption{\mbox{R-band} surface brightness profiles measured along the axis of the trail and averaged over $10$ pixels perpendicular to the trail and 2 pixels along the trail ($1.19\times0.24$~arcsec). The dark thin line is the profile measured in the averaged data (\emph{top panel} of Fig.~\ref{Fig_AvgImage}), which clearly shows the nucleus (upwards arrow) and the four fragments (downwards arrows). The underlying light thick line is the profile of a synthetic trail that was subtracted from the averaged data (cf.~\emph{bottom panel} of Fig.~\ref{Fig_AvgImage}). \emph{Top panel} covers the same portion of the trail axis as the \emph{top panel} of Fig.~\ref{Fig_AvgImage}. \emph{Bottom panel} shows a close-up view of the region occupied by the main nucleus and fragments, consistent with the $3\times$ zoom presented in the \emph{bottom panel} of Fig.~\ref{Fig_AvgImage}.\label{Fig_TrailProfiles}}
\end{figure}

The average image of \mbox{P/2012 F5} (Fig.~\ref{Fig_AvgImage}) shows the star-like nucleus and the broader trail (Fig.~\ref{Fig_Profiles}), consistent with previous observations \citep{Ste12,Mor12,Nov14}, but also reveals for the first time at least four fragments of the main nucleus. The fragments are embedded in the trail and separated from the nucleus by $3.1$, $12.6$, $22.1$, and $29.3$~arcsec (Fig.~\ref{Fig_TrailProfiles}). The nearest two fragments are $\sim5$ and $4.2\pm0.1$~mag fainter than the average brightness of the main nucleus. Assuming that they are inactive, discrete objects, their magnitudes correspond to the equivalent diameters of $180$ and $250$~m, respectively. The two more distant components are also $\sim5$~mag fainter than the main nucleus, but because they may be elongated in the same direction as the trail, it is highly uncertain whether they are discrete objects or not. Future observations are needed to detect the relative motion and morphological evolution of the discovered fragments, and in this way establish their nature and separation parameters.

\section{Analysis}\label{Sec_An}

To date, the nucleus rotation period has been measured for eight active asteroids \citep[see][]{Jew15} and three asteroids (\mbox{P/2010 A2}, \mbox{P/2013 R3}, and \mbox{P/2012 F5}) were observed to experience nucleus fragmentation. Of all these objects \mbox{P/2012 F5} is the only one with both properties robustly measured. It is also the second active asteroid ejecting dust in impulsive manner, after (596)~Scheila, for which the nucleus rotation period is known. However, unlike for the slowly rotating Scheila, the spin of \mbox{P/2012 F5} is fast enough to explain the sudden ejection of dust and fragments by rotational instability.

Let us approximate the nucleus of \mbox{P/2012 F5} as a prolate spheroid with a mass $M$, length of the long semi-axis $a$, and long-to-short axis ratio $f$. The gravitational acceleration above the tip of the long axis is:
\begin{equation}\label{Eq_gr}
g(r) = \frac{GM}{r^2}\Bigg(\frac{3r^2}{a^2\phi^2}\,\Big[\frac{r}{2a\phi}\,\ln\Big(\frac{r+a\phi}{r-a\phi}\Big)-1\Big]\Bigg),
\end{equation}
where $G = 6.67384 \times 10^{-11}$~m$^3$~kg$^{-1}$~s$^{-2}$ is the gravitational constant, $\phi = \sqrt{1-1/f^2}$ is a function of the long-to-short axis ratio, and $r \ge a$ is the distance from the gravity center. Note that the term in the outer brackets of the above equation approaches unity (regardless of $r$ or $a$) for a sphere, for which $f=1$ and thus $\phi=0$. Using Equation~\ref{Eq_gr} we calculate the escape speed at the tip of the long axis, assuming that the long axis is always pointed toward the escaping object. We obtain:
\begin{equation}\label{Eq_ve}
v_\mathrm{e} = \sqrt{2\int_a^\infty\!g(r)\,\mathrm{d}r} = \sqrt{\frac{2GM}{a}\,\Bigg(\frac{3}{2}\,\frac{(\phi^2-1)\,\,\mathrm{arctanh}\,(\phi) + \phi}{\phi^3}\Bigg)},
\end{equation}
where the term in the outer brackets again approaches unity for a sphere. In fact a lower ejection speed may be sufficient to break free from the gravitational attraction of the nucleus and therefore the derived escape speed is an upper limit. This is because it is enough to reach the nucleus Hill sphere instead of infinity and also because the gravitational attraction becomes smaller when the long axis is eventually pointed away from the escaping object due to nucleus rotation. Assuming that the nucleus rotates in a simple mode around the short axis, let us now calculate the rotation period $P$ for which the rotational speed at the tip of the log axis, $v_\mathrm{r} = 2\,\pi\,a/P$, is equal to the escape speed $v_\mathrm{e}$ given by the above equation. Keeping in mind that a prolate spheroid with a density $\rho$ has a mass $M = 4/3\,\pi\,\rho\,a^3/f^2$, we obtain:
\begin{equation}\label{Eq_Pe}
P_\mathrm{e} = \sqrt{\frac{3\pi}{2G\rho}\,\Bigg(\frac{2}{3}\,\frac{\phi^3}{(1-\phi^2)[(\phi^2-1)\,\,\mathrm{arctanh}\,(\phi) + \phi]}\Bigg)},
\end{equation}
where the term in the outer brackets once again approaches unity for a sphere. Assuming that the bulk density of \mbox{P/2012 F5} is in the range $1000$--$2000$~kg~m$^{-3}$ (expected for \mbox{C-type} asteroids), we find from the above equation that the object's rotation is fast enough to launch fragments to infinity (i.e., $P_\mathrm{e} \ge P = 3.24$~hr), if the axis ratio $f$ is at least $1.49$--$2.21$, in agreement with the measured lower limit $f \ge 1.204$.

Now let us balance the (centripetal) gravitational acceleration $g(r)$ from Equation~\ref{Eq_gr} with the apparent centrifugal acceleration $u(r) = 4\pi^2r/P^2$, again substituting mass $M = 4/3\,\pi\,\rho\,a^3/f^2$. Assuming that the nucleus is strengthless, we request that the two accelerations cancel out at the tip of the long axis, $g(a) = u(a)$, which allows us to find the critical rotation period for rotational breakup:
\begin{equation}\label{Eq_Pc}
P_\mathrm{c} = \sqrt{\frac{3\pi}{G\rho}\,\Bigg(\frac{2}{3}\,\frac{\phi^3}{(1-\phi^2)[\ln\big(\frac{1+\phi}{1-\phi}\big) - 2\phi]}\Bigg)}.
\end{equation}
Note that the term in the outer brackets also this time approaches unity for a sphere, and is very close to $f$ in a wide range of axis ratios (for $f$ between $1.0$ and $3.0$ the approximation is good to better than 5\%). Using the above equation we find that for the same combination of density and axis ratio as before, the object's self-gravity becomes insufficient to ensure rotational stability if the rotation period is shorter than $3.93$--$3.42$~hr. The fact that \mbox{P/2012 F5} spins faster than this limit is consistent with the action of weak cohesive forces increasing the nucleus stability. The value of the cohesive strength cannot be calculated (given the unknown cross sections of the planes of failure), but for an object with the same properties as before (except for zero strength) and the volume-equivalent diameter of $1.77$~km as estimated for \mbox{P/2012 F5}, the strength needs to be greater than $\sim30$~Pa to prevent the model nucleus from splitting in half \citep[cf. e.g.][]{Dav01,Hir14}. This value is consistent with the minimum cohesive strength of $64$~Pa obtained for the rapidly rotating rubble-pile asteroid (29075) and other estimations \citep[see][and ref. therein]{Roz14}.

Although our result does not uniquely establish the fragmentation of \mbox{P/2012 F5} as caused by the rapid nucleus spin, it is evident that the conditions for material failure on the surface and fragment ejection at the escape speed are simultaneously satisfied for this object for a plausible range of (unknown) parameters, making this scenario certainly possible. However, one should keep in mind that the probability of a hypervelocity collision is obviously the same for fast and slow rotators, and therefore the possibility of an impact-generated fragmentation cannot be completely ruled out for \mbox{P/2012 F5} at this time.

\section{Discussion}\label{Sec_Disc}

While \mbox{P/2012 F5} has the fastest known spin among the active asteroids, it is not the only rapid rotator in this group. The other two objects in the Main Belt are: \mbox{133P/Elst}-Pizarro --- the prototype active asteroid with a confirmed protracted and recurrent mass loss \citep[e.g.][]{Hsi10}, and (62412) with an uncertain type of mass loss \citep{She15}, which have the rotation periods of $3.471$ and $3.33$~hr, respectively \citep{Hsi04,She15}. There is also the sun-skirting active asteroid (3200)~Phaethon, which shows a recurrent mass loss near the Sun \citep{Jew10b,Li13,Jew13a} and has a short rotation period of $3.603$~hr \citep{Ans14}. The other five active asteroids with known spin rates are slow rotators displaying various types of mass loss. For example, \mbox{176P/LINEAR} has a rotation period of $22.23$~hr and shows a protracted activity \citep{Hsi11}, while the already-mentioned (596)~Scheila has a rotation period of $15.848$~hr \citep{War06} and ejected dust impulsively due to an impact. These measurements show the existence of a population of rapid rotators among the active asteroids, but curiously, there is no link between the spin rate and the type of mass loss. Therefore, an intriguing hypothesis can be postulated that different physical mechanisms triggering the mass loss of active asteroids are in fact not discriminated by the common classification of these objects according to the type of observed activity, but instead, by the rotational stability of the nucleus. Mass loss demonstrated by the rapidly rotating active asteroids may have a general origin in their rotational instability, whereas another mechanism, such as hypervelocity impact, triggers the activity of the slowly rotating objects. Each mechanism can directly produce and launch dust and fragments, but also excavate ice which may subsequently sublimate, generating the various types of mass loss displayed by the active asteroids.

In particular, the mass loss from 133P is best explained by a combination of the rapid nucleus spin and seasonal ice sublimation \citep{Jew14b}, while the activity of (62412), though possibly driven by sublimating ice, might have been initiated by ice-excavating change of the nucleus shape caused by the rapid spin \citep{She15}. The rapid spin may also support surface disintegration and ejection of material observed in the sun-skirting Phaethon \citep{Jew10b}. Likewise, the \mbox{X-shaped} \mbox{P/2010 A2} and the multi-tailed \mbox{P/2013 P5} are suspected of rapid nucleus spin, leading to a partial fragmentation and multiple landslides, respectively \citep{Aga13,Jew13b}. Also the total fragmentation of \mbox{P/2013 R3} might have resulted from rotational instability of the nucleus \citep{Jew14a,Hir14}. In contrast, the short-lasting ejecta plumes emanating from the slowly rotating Scheila were explained as a direct result of a hypervelocity impact \citep[e.g.][]{Ish11b}, whereas the slow spin and protracted dust emission of 176P make it so far the best candidate for the activity driven by sublimation of ice earlier excavated by an impact --- the original mechanism proposed to explain the mass loss from the first discovered active asteroids \citep{Hsi06}.

\acknowledgments

We thank Masatoshi Hirabayashi and the anonymous referee for comments. M.~D. is grateful for support from the National Science Centre of Poland through a FUGA Fellowship grant 2014/12/S/ST9/00426. D.~J. appreciates support from NASA's Solar System Observations program. We thank the staff of the W. M. Keck Observatory for assistance and are indebted to Caltech Optical Observatories for allocating Keck~II time for this program.

{\it Facilities:} \facility{Keck:II (DEIMOS)}.

\clearpage

\clearpage

\clearpage

\clearpage

\end{document}